\documentclass[traditabstract]{aa}

\usepackage{graphicx, times, txfonts, float, rotating, color, url, lscape, bm, placeins} 

\newcommand{\less}{\raisebox{-1.1mm}{$\stackrel{<}{\sim}$}} 
\newcommand{\more}{\raisebox{-1.1mm}{$\stackrel{>}{\sim}$}} 
\newcommand{\msol}{\mbox{M$_{\odot}$}}

\newcommand{\lsol}{\mbox{L$_{\odot}$}}

\newcommand{\ks}{km s$^{-1}$}

\newcommand{\G}{{\it Gaia}}

\begin{document} 
 
\title{The spectral energy distributions of very long-period Cepheids in the Milky Way, the Magellanic Clouds, M31, and M33
}  
 
\author{
  M.~A.~T.~Groenewegen
}

\institute{ 
Koninklijke Sterrenwacht van Belgi\"e, Ringlaan 3, B--1180 Brussels, Belgium \\ \email{martin.groenewegen@oma.be}
} 
 
\date{received: ** 2025, accepted: * 2025} 


\titlerunning{The SEDs of very long-period cepheids in the Milky Way, the Magellanic Clouds, M31 and M33} 
 
\abstract
{
  The spectral energy distributions (SEDs) of 20 Milky Way (MW), 9 Large Magellanic Cloud (LMC), 7 Small Magellanic Cloud (SMC),
  12 M31,
  and 7 M33  (classical) Cepheids with periods longer than 50 days were constructed using
photometric data from the literature and fitted with model atmospheres with the aim of identifying objects with an infrared excess.
The SEDs were fitted with stellar photosphere models to derive the best-fitting luminosity and effective temperature; 
a dust component was added when required.
The distance and reddening values were taken from the literature.
WISE and IRAC images were inspected to verify whether potential excess emission was related to the central objects.

Only one star with a significant infrared (IR) excess was found in the LMC and none in the SMC, M31, and M33,
contrary to earlier work on the MW suggesting that IR excess may be more prominent in MW Cepheids than
in the Magellanic Clouds. One additional object in the MW was found to have an IR excess, but it is unclear whether
it is a classical Cepheid or a type-{\sc ii} Cepheid.

The stars were plotted in a Hertzsprung--Russell diagram (HRD) and compared
to evolutionary tracks for CCs and to theoretical instability strips.
For the large majority of stars, the position in the HRD is consistent with the instability strip.
For stars in the MW uncertainties in the distance and reddening can significantly change their position in the HRD.
}


\maketitle

\section{Introduction}
\label{S:Int}

Classical Cepheids (CCs) are important standard candles because they are bright and provide a link
between the distance scale in the nearby universe and that further out via those galaxies that contain both Cepheids and SNIa 
(see \citealt{Riess2022} and \citealt{Murakami2023} for a determination of the Hubble constant to 1.0~\ks\ precision or better).
Typically, the period-luminosity (PL) relations of CCs that are at the core of the distance determinations are derived
in particular photometric filters ($V, I, K$) or combinations of filters that are designed to be
reddening independent, called the Wesenheit functions \citep{Madore82}, for example using combinations of ($V,I$) or
($J,K$), or the combination used by the SH0ES team (F555W, F814W, and F160W HST filters; see \citealt{Riess2022}).

On the other hand, the bolometric magnitude or luminosity is a fundamental quantity of stars as it is the
output of stellar evolution models and the input to CC pulsation models.
This is the continuation of a series of papers that construct and analyse the spectral energy distributions (SEDs) of CCs.
In \citet{GrSED} (hereafter G20) the SEDs of 477 Galactic CCs were constructed and
fitted with model atmospheres (and a dust component when required). For an adopted distance (from {\it Gaia} DR2 at that time), 
reddening these fits resulted in a
best-fitting bolometric luminosity ($L$) and the photometrically derived effective temperature ($T_{\rm eff}$).
This allowed the derivation of
period-radius ($PR$) and period-luminosity ($PL$) relations, the construction of the Hertzsprung--Russell diagram (HRD), and a comparison to
theoretical instability strips (ISs).
This sample was further studied in \citet{GrFWG}, where the relation was investigated between
the bolometric absolute magnitude and the flux-weighted gravity (FWG); this is known as
the flux-weighted gravity-luminosity relation (FWGLR).

In \citet{GrLub} 77 Small Magellanic Cloud (SMC) and 142 Large Magellanic Cloud (LMC) CCs were studied along similar lines.
The advantage of using the Magellanic Clouds (MCs) is that accurate and independently derived mean distances are available based on
the analysis of samples of eclipsing binaries \citep{Pietrzynski19, Graczyk20}.

Interestingly, in the latter study, only one case was found where there was evidence of an infrared excess, namely the longest period object in the LMC.
This is in contrast to Galactic CCs where near-IR (NIR) and mid-IR (MIR) excess is known to exist, revealed
for example via direct interferometric observations
in the optical or NIR (e.g. \citealt{Kervella2006, Merand2006, GalVISIR, Nardetto16, Hocde25}),
modelling with the SPIPS code (e.g. \citealt{Breitfelder16, Trahin19, Trahin21}, and \citealt{Gallenne17} for the LMC) 
and was also found in modelling of the SEDs of Galactic CCs (\citealt{Gallenne13b}, G20).

This raises the question of whether this apparent difference in the presence of IR excess could be related to metallicity.
To investigate this further, a complete sample of long-period Cepheids (periods longer than 50~days; see below)
is studied in this paper, in the MW, the MCs, and M31 and M33.
This study is connected to the class of ultra long-period (ULP) Cepheids, a term introduced by \citet{Bird09} as
fundamental mode (FU) Cepheids with periods longer than 80~days
(see reviews by \citealt{Musella21} and \citealt{Musella22} specifically on ULPs).

The paper is structured as follows.
In Section~\ref{S-Sam} the sample of Cepheids is introduced, while 
Section~\ref{S:PDM} introduces the photometry that is used, the distances used, and how the
modelling of the SED was done.
Section~\ref{S:Res} discusses several results, in particular the location of the objects in the HRD,
the presence of infrared excess, the PR and PL relations, and models with alternative distances or reddenings.
A brief discussion and summary concludes the paper in Sect.~\ref{S:Dis}.

\section{Sample} 
\label{S-Sam}

For this paper a sample of 55 Cepheids was studied.
In particular the sample is compiled from the following:

\begin{itemize}

\item Galactic Cepheids from \citet{Pietrukowicz21}\footnote{Version dated September 17, 2022
  \url{https://www.astrouw.edu.pl/ogle/ogle4/OCVS/allGalCep.listID}} which contains 3666 CCs.
The longest period listed there is S Vul with a period of 68.65~d, clearly shorter than the classical limit of 80 days
for ULPs.
An (arbitrary) lower limit of 50 days is used, which results in nine objects.

\item SMC and LMC CCs from the OGLE-{\sc iv} catalogue \citep{Soszynski19}, resulting in six and eight objects, respectively, with
  periods longer than 50 days.

\item From the {\it Gaia} DR3 {\tt vari\_cepheid} table all CCs with a period longer than 50 days and type DCEP were selected,
  for a total of 53 objects \citep{RipepiDR3Cep22,GaiaDR3Vallenari22,GC2016a}.  

\end{itemize}

\noindent
All of the sources from \citet{Pietrukowicz21} are in the {\tt vari\_cepheid} table, except OGLE-GD-CEP-1505.
It is listed there, but classified as a Type-{\sc ii} Cepheid (T2C) of the RV Tau class.
It was kept in the sample as our analysis may shed light on its nature.
All of the sources from \citet{Soszynski19}    are in the {\tt vari\_cepheid} table, except OGLE-LMC-CEP-4689.
This is the well-known variable HV 2827, listed in the \G\ main catalogue, but not in the  \G\ Cepheid and {\tt vari\_summary} tables. The source is kept.
Thirty-one sources from the  {\tt vari\_cepheid} table are not in the samples from \citet{Pietrukowicz21} and  \citet{Soszynski19}.

The 55 sources were matched with the SIMBAD database to obtain additional names and identifiers.
Twelve objects are likely members of M31  and seven are likely members of M33.
The remaining 20 appear to be in the Milky Way.
Five of them are in the direction of the Galactic Bulge  and four of these have been classified as T2C by the OGLE team.

Basic information of the 55 stars are compiled in Table~\ref{Tab:Sample}.
All LMC objects except LMC-Dachs2-24, and all SMC objects except SMC-Dachs3-5 and SMC-CEP-1977 were studied by \citet{GrLub}, while
S Vul and GY Sge were studied in G20. However, the analysis of the SEDs was repeated here independently.
It should be noted that there are known CCs in other galaxies with periods longer than 50 days
(see e.g. \citealt{Musella22}), but they are not included in the {\it Gaia} {\tt vari\_cepheid} table. 

As for some sources there is a possible confusion about whether they are CCs or T2C of the RV Tau type,
Cols. 8 and 9 give the predicted luminosity for the two classes based on the LMC PL relations of \citet{GrLub}
and \citet{GrJu17b}, respectively.
Typically, these luminosities differ by a factor of 20-30 for periods in the range 50-200~days, implying
changes in distance by a factor of 5 to
`convert' a T2C into a CC, or vice versa, purely based on consistency with a PL relation.

\section{Photometry, distance, masses, and modelling}
\label{S:PDM}

\subsection{Photometry}
\label{SS-phot}

The SEDs were constructed using photometry retrieved mostly, but not exclusively, 
via the VizieR web-interface\footnote{\url{http://vizier.u-strasbg.fr/viz-bin/VizieR}}.
Table~\ref{Tab-Phot} lists the filters and references to the photometry that were considered.
An additional reason for not considering known CCs with periods longer than 50 days in more distant galaxies is that there are
less (and less accurate) MIR data available, which are needed to detect an IR excess, and the problem of contamination or
blending as a given beam size or aperture corresponds to a larger physical size (as indeed turns out to be the case for some objects, see Sect.~4.3).
The data contain single-epoch observations (typically from GALEX and Akari) but whenever possible values
at mean light were taken or multiple data points were averaged.

\subsection{Distance and geometric correction}
\label{SS-dist}

For the LMC, M31, and M33, mean distances from the literature were adopted and a geometric correction was applied to
  correct for the fact that the sources are to first order located in an inclined disc, following \citet{Grocholski07}.
The depth effect in the SMC is considerable (e.g. \citealt{Ripepi17}), and all SMC sources were adopted to be at the mean distance.
For the Galactic Bulge region\footnote{Defined as the region with $266 < R.A. < 270\degr$ and $-33 < \delta  < -28\degr$ and comprising the
  four sources with BLG-T2CEP in their names and the source marked GDR404357.} the distance from the GRAVITY experiment was adopted \citep{Gravity22}.
For the remainder of the MW sources the geometric distance from \citet{BJ21} was adopted.
Details and references are listed in Table~\ref{Tab:Distance}.
The distances to the individual sources are given in Table~\ref{Tab-Res}.

The 3D reddening maps of \citet{Lallement22} and \citet{Vergely22} were used\footnote{\url{https://explore-platform.eu}} to obtain the $A_{\rm V}$
in a given direction as well as the distance to which this reddening refers.
For the sources in M31 and M33 a value of  $A_{\rm V}= 0.17$ was adopted which is the average value from the 3D reddening map at the
largest available distance in the direction of those galaxies.
For the MCs the reddening map of \citet{Skowron21} was used and the $E(V-I)$ value in the map closest to the source is taken.
The visual extinction was then taken as $A_{\rm V}= 3.1 \cdot E(V-I) /1.318$, following \citet{Skowron21}.  
The reddenings to the individual sources are given in Table~\ref{Tab-Res}.

\begin{table*}
  \centering

\caption{Adopted distances}
\begin{tabular}{lll} \hline \hline 
  Region & Distance model & References \\
         & (kpc) & \\
  \hline   
  M33 & 840 $\pm$ 11 + geometric correction ($i= 57$\degr, PA= 22.5\degr) & \citet{Breuval23}, \citet{Kourkchi20} \\   
  M31 & 761 $\pm$ 11 + geometric correction ($i= 70$\degr, PA= 43\degr)   & \citet{Li2021}, \citet{Dalcanton12} \\
  SMC &  62.44 $\pm$ 0.94                                                     &  \citet{Graczyk20} \\
  LMC &  49.59 $\pm$ 0.55 + geometric correction ($i= 25$\degr, PA= 132\degr) &  \citet{Pietrzynski19}, \citet{Riess19} \\
  BUL &   8.28 $\pm$ 0.03  & \citet{Gravity22, Gravity21}  \\   
  MW  & see text           & - \\
\hline
\end{tabular} 
\label{Tab:Distance}
\end{table*}

\subsection{Modelling}
\label{SS-model}

The SEDs are fitted with the code
More of DUSTY (MoD, \cite{Gr_MOD})\footnote{\url{http://homepage.oma.be/marting/codes.html}}, 
which uses a slightly updated and modified version of the DUSTY dust radiative transfer (RT) code \citep{Ivezic_D} as
a subroutine within a minimisation code. The dust optical depth is initially set to zero.
In that case the inputs to the model are the distance, reddening, and a model atmosphere.
The cases where an infrared (IR) excess may be present are discussed in Sect.~\ref{S:dust}.

The MARCS model atmospheres are used as input \citep{Gustafsson_MARCS} for $\log g= 1.5$ and for adopting canonical metallicities
of $-0.50$ and $-0.75$ dex for the LMC and SMC stars, and $+0.00$ for M31, M33, the Bulge, and the
MW sources\footnote{Metallicity gradients in M31 and M33 are not considered (see e.g. \citealt{Li2025}).}.
The model grid was available at 250~K intervals for the effective temperature 
range of interest, and adjacent model atmospheres were used to interpolate models at 125~K intervals, 
which more closely reflects the accuracy in $T_{\rm eff}$ that can be achieved.
For every model atmosphere (that is, $T_{\rm eff}$) a best-fitting luminosity (with its [internal] error bar, based on the covariance
matrix) is derived with the corresponding reduced $\chi^2$ ($\chi_{\rm r}^2$) of the fit. 
The model with the lowest $\chi_{\rm r}^2$ then gives the best-fitting effective temperature. 
Considering models within a certain range above this minimum $\chi_{\rm r}^2$ then gives the estimated error in the 
effective temperature and luminosity. For the luminosity this error is added in quadrature to the internal error in luminosity.

In the model fitting procedure photometric outliers were excluded in the following way.
The photometric error bar for each data point was added in quadrature to 1.4826 $\cdot$ median absolute deviation (MAD) of
the residuals in the fit to give the equivalent of 1$\sigma$ in a Gaussian distribution.
If the absolute difference between the model and observations was larger than 4$\sigma$, the point was flagged and
plotted with an error bar of 3.0~mag so that it was still identified, but had no influence on the fitting.
The model grid over temperatures was run again, and the clipping procedure repeated. Then the run over effective temperatures was repeated a last time.
The best-fitting effective temperature and luminosity with error bars are listed in Table~\ref{Tab-Res}.

\begin{figure}
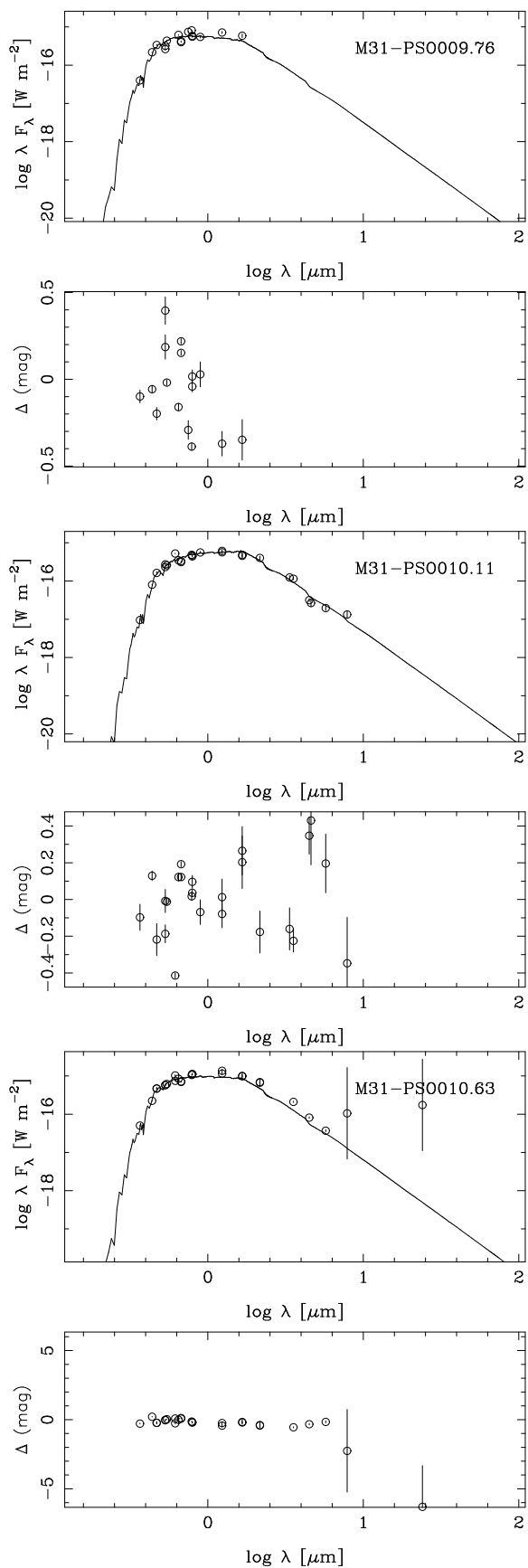


\begin{minipage}{0.41\textwidth}
\resizebox{\hsize}{!}{\includegraphics{M31-PSO009.76_sed.ps}}
\end{minipage}
\begin{minipage}{0.41\textwidth}
\resizebox{\hsize}{!}{\includegraphics{M31-PSO010.11_sed.ps}}
\end{minipage}
\begin{minipage}{0.41\textwidth}
\resizebox{\hsize}{!}{\includegraphics{M31-PSO010.63_sed.ps}}
\end{minipage}

\caption{Examples of best-fitting models assuming no dust. The upper panels show the observations (with error bars) and the model.
  The lower panel shows the residuals. Outliers that have been clipped are plotted with an (arbitrary) error bar of 3.0~mag.
}
\label{Fig:sed1}
\end{figure}

\section{Results}
\label{S:Res}

\subsection{General}

Figure~\ref{Fig:sed1} shows some best fits without considering dust. This illustrates the quality of the modelling with the residual
(model minus observations) in the bottom part of each panel\footnote{The complete set of SEDs is
  available at \protect\url{https://doi.org/10.5281/zenodo.15422721}}.

\subsection{Hertzsprung--Russell diagram}

\begin{figure*}
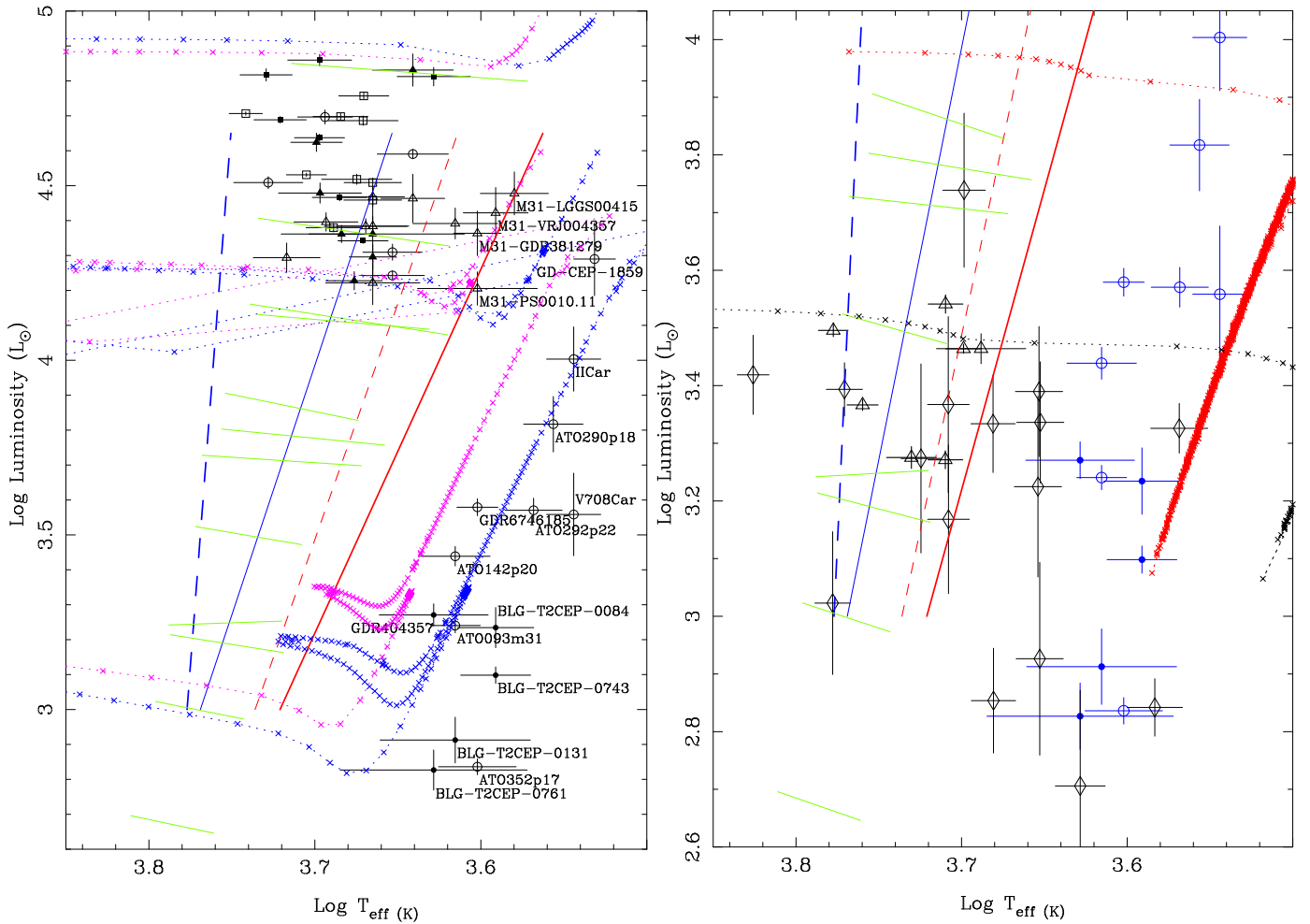

  \centering

\begin{minipage}{0.49\textwidth}
\resizebox{\hsize}{!}{\includegraphics{HRD_Lum_Teff.ps}}
\end{minipage}
\begin{minipage}{0.49\textwidth}
\resizebox{\hsize}{!}{\includegraphics{HRD_Lum_Teff_T2C.ps}}
\end{minipage}

\caption{Hertzsprung--Russell diagram.
The left panel presents an overview while the red panel focusses on the T2Cs.
The symbols are follows: filled squares (SMC), open squares (LMC), open triangles (M31),
filled triangles (M33), filled circles (Galactic Bulge), and open circles (MW).
Stars located outside the bulk of objects are identified.
The blue and red lines indicate the blue and red edge of the IS of CCs.
The results from \citet{DeSomma21} are plotted for $Z = 0.03$ (thick solid lines)
and $Z = 0.004$ models (thinner dashed lines), for their type A mass-luminosity relation.
The green lines indicate evolutionary models from \citet{Anderson16} (see text for details).
The blue and pink crosses indicate MIST evolutionary tracks for $v/v_{\rm crit}=0.4$
for 5.0, 10, and 17~\msol\ (and [Fe/H]= 0.0~dex)
and 4.8,  9, and 15~\msol\ (and [Fe/H]= $-0.50$~dex), respectively,  plotted at 10$^4$ year intervals
\citep{Dotter16,Choi16,Paxton15,Paxton13,Paxton11}. The first crossing of the IS is also visible, except for the highest mass tracks.
\\
In the right panel the objects in the sample are plotted in blue, without names and
the MIST evolutionary tracks are not shown.
Instead, the black and red crosses indicate the evolutionary tracks of the 1.0 and 2.5~\msol\ solar metallicity stars,
respectively, including the post-AGB phase, plotted at 10$^3$ year intervals (from \citealt{VW94}).  
For comparison, the black triangles and spades indicate T2C with periods longer than 50~days in the MCs (from \citealt{GrJu17b})
and the MW (from \citealt{BodiKiss19}), respectively.
GD-CEP-1505 is located outside both plots at $\log T_{\rm eff} \sim 3.48$ and $\log L \sim 2.18$.
}
\label{Fig:HRD}
\end{figure*}

Figure~\ref{Fig:HRD} shows the HRD together with sets of evolutionary tracks and the ISs of CCs in two panels.
Objects from the sample are plotted as filled squares (SMC), open squares (LMC), open triangles (M31),
filled triangles (M33), filled circles (Galactic Bulge), and open circles (MW).
Stars located outside the bulk of objects are plotted with error bars and some are labelled as well.
The red and blue edges of the IS of CCs are plotted for Z= 0.015 and 0.004 \citep{DeSomma21}.
The near horizontal green lines indicate the evolutionary tracks of CCs for $Z = 0.014$ and average initial rotation rate
$\omega_{\rm ini} = 0.5$ from \citet{Anderson16}.  Increasing in luminosity are tracks for initial mass
(the number of the crossing through the IS): 4 (1), 5 (1), 5 (2), 5 (3), 7 (1), 7 (2), 7 (3), 9 (1), 9 (2), 9 (3), 12 (1), and 15 \msol\ (1).
The blue and pink crosses in the left panel indicate the MIST evolutionary tracks 
for 5.0, 10, and 17~\msol\ (for [Fe/H]= 0.0~dex)
and 4.8,  9, and 15~\msol\ (for [Fe/H]= $-0.50$~dex), respectively,  plotted at 10$^4$ year intervals
\citep{Dotter16,Choi16,Paxton15,Paxton13,Paxton11}.
The 4.8 and 5.0~\msol\ tracks are the lowest mass ones with blue loops that reach the IS of CCs.

The right panel focusses on the T2C, and the black and red crosses respectively indicate the evolutionary tracks
of 1.0 and 2.5~\msol\ solar metallicity stars, including the post-AGB phase, plotted at 10$^3$ year
intervals (from \citealt{VW94}).  
In addition, triangles and diamonds indicate T2C with periods longer than 50~days in the MCs (from \citealt{GrJu17b})
and the MW (from \citealt{BodiKiss19}), respectively.

In the left panel there is a clear separation between objects whose location is consistent with the IS of CCs, and
those that are clearly cooler, and in almost all cases, are significantly less luminous.
Most of them have been classified as T2C in the literature.
It is noted that 4 of the 12 CCs in M31 are close to the red edge of the IS for $Z= 0.03$.
Based on the MIST and \citet{Anderson16} evolutionary models the CCs have masses in the $\sim$10-15~\msol\ range.

The right panel, which focusses on the T2C region shows that the location of the stars in the sample is different
from the known MCs and MW T2Cs with periods longer than 50~days from \citet{GrJu17b} and \citet{BodiKiss19}.
The selection was very different in the sense that the latter studies started from samples of T2C, while
the stars in this sample were initially believed to be largely CCs.
Six objects have luminosities that lie above or close to the post-AGB track for a 1~\msol\ track, but seven do not.
These objects are most probably post-RGB objects (see \citealt{Kamath2016}), although they are not dusty.
Among the potential T2Cs only II Car shows some evidence of the presence of dust (see Sect.~\ref{S:dust}), all the
others are well fit by a stellar atmosphere and do not show the characteristic disc or shell IR signature
in their SEDs.

\subsection{Infrared excess}
\label{S:dust}

The default assumption in the modelling was that there is no IR excess and the SEDs can be modelled by a stellar atmosphere.
However, NIR and MIR excess are known to exist in Galactic CCs, for example direct interferometric observations in the optical or NIR
and other methods (see references quoted in the Introduction),
and one possibility to explain the IR excess is through dust emission.
  For T2Cs dust emission is a very plausible explanation as the long period T2C are associated with the RV Tau variability class that are generally believed to be
  in the post-AGB evolutionary phase (e.g. \citealt{Manick18}).

In a next step, models were run with the dust optical depth (${\tau}_{\rm d}$, at 0.5~$\mu m$) and dust temperature at the inner radius ($T_{\rm d}$)
as additional fit parameters. The dust shell was assumed to be spherically symmetric. Models with different initial guesses were run
($T_{\rm d}$ starting from 250, 400, 600, 800, 1000, and 1500~K; ${\tau}_{\rm d}$ starting from 0.1, 0.3, 0.6, and 1).
The Bayesian information criterion (BIC; \citealt{Schwarz1978}) was used as a first check to determine
whether a model with dust fitted the SED better than a model atmosphere.
However, some flexibility in a strict application was needed as some seemingly better models converged to $T_{\rm d}$ values higher than
the effective temperature, or had an error bar on $T_{\rm d}$ of the same order as $T_{\rm d}$.
In addition, the initial set of models was run for an effective temperature that resulted from the models without dust.
A model with dust will increase the flux in the infrared due to emission but it will also absorb radiation in the optical,
and therefore the best-fitting effective temperature will likely become higher.
  It should be pointed out that for a few objects (M31-PSO009.76, shown in Fig.~\ref{Fig:sed1}, M33-013331, M33-V00021, and  M33-013405)
  there is no photometry available beyond the NIR. As the contrast between the emission by dust and the stellar photosphere increases
  with wavelength this is not ideal to detect any excess emission; the absence of proof for infrared excess is not the proof of absence.

\begin{figure*}
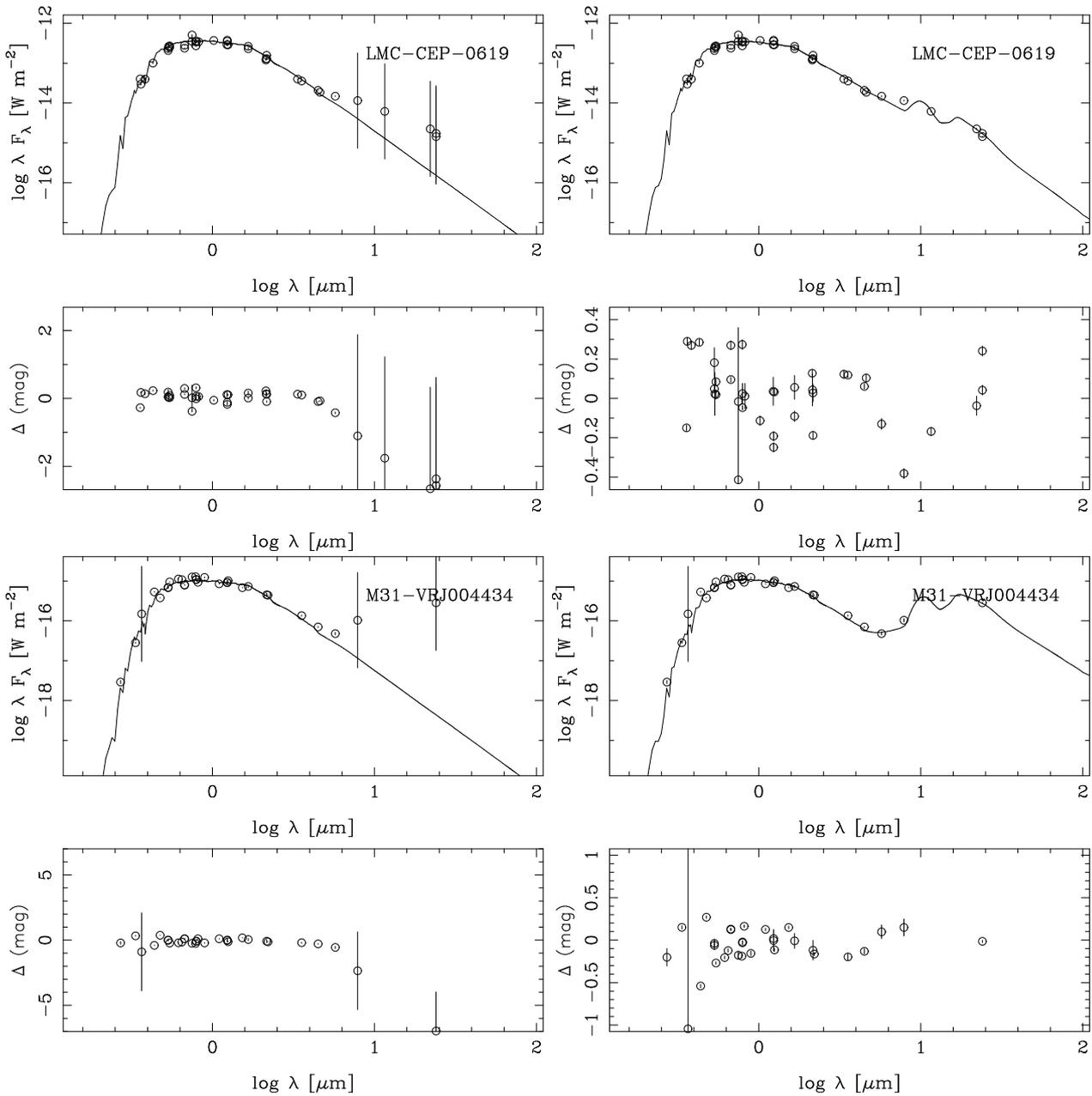


\begin{minipage}{0.45\textwidth}
\resizebox{\hsize}{!}{\includegraphics{LMC-CEP-0619_sed.ps}}
\end{minipage}
\begin{minipage}{0.45\textwidth}
\resizebox{\hsize}{!}{\includegraphics{LMC-CEP-0619_sed_dusty.ps}}
\end{minipage}

\begin{minipage}{0.45\textwidth}
\resizebox{\hsize}{!}{\includegraphics{M31-VRJ004434_sed.ps}}
\end{minipage}
\begin{minipage}{0.45\textwidth}
\resizebox{\hsize}{!}{\includegraphics{M31-VRJ004434_sed_dusty.ps}}
\end{minipage}

\caption{Examples of best-fitting models assuming dust (right) compared to no dust (left panels).
  We note the difference in the range of the ordinate in the left and right bottom panels.
  The other six objects are shown in Fig.~\ref{Fig:seddusty}.
  In the models without dust some photometric points are considered outliers and are plotted with a large error bar, instead of omitting them.
}
\label{Fig:sed2}
\end{figure*}

Based on these considerations, additional models were run for 9 stars out of the 55 in the sample, where the effective
temperature was allowed to vary as well.
For one star the results were not deemed conclusive  (M33-013305) and the best-fitting models including dust for eight stars
are listed in Table~\ref{Tab-ResDust}, one star each in the MW, LMC, and M33, and five in M31.
Figure~\ref{Fig:sed2} compares the best-fitting models with and without dust for two objects, while the six others are shown in Fig.~\ref{Fig:seddusty}. 
Table~\ref{Tab-ResDust} includes the BIC of models under three model assumptions; the SEDs fitted including a dust component,
a model without dust and where photometric outliers were removed (the model in Table~\ref{Tab-Res}), and 
a model without dust and including all data points.
The last model clearly represents the worst fit.
In six cases the BIC of the model including dust is lower than the model without the outliers.
Comparing the reduced $\chi^2$ (Col.~4 in Table~\ref{Tab-ResDust} to Col.~6 in Table~\ref{Tab-Res}) this is only the case for two objects.

To further investigate the nature of the IR excess and to check the possibility that the IR excess is not
associated with the CCs but
related to diffuse background, the emission images from the ALLWISE survey\footnote{\url{https://irsa.ipac.caltech.edu/applications/wise/}}
and the Spitzer Enhanced Imaging Products (SEIP)\footnote{\url{https://irsa.ipac.caltech.edu/data/SPITZER/Enhanced/SEIP/}} were inspected.
Figure~\ref{Fig:WISE} and Figure~\ref{Fig:IRAC} show images in the WISE W1 and W3, and IRAC 1 and 4 filters, respectively, around selected CCs.
The two objects in the top rows are different examples of stars without IR excess that are located in empty fields.
The other panels show the eight stars with an IR excess in the SEDs from Table~\ref{Tab-ResDust}.
Except for II Car and LMC-CEP-0619, the emission of the stars at the longest WISE3 and IRAC4 wavelengths (when detected) seems not be
clearly associated with the central star, but appears to be diffuse emission or possibly partly blended.
  In the cases of M31-PSO010.63, M31-GDR369249, M31-VRJ004357,  M31-PSO011.09, M31-VRJ004434, 
  and M33-013312 the correct model is the standard model without dust and not considering the photometric outliers because
  these are very likely not associated with the objects.


\begin{table*}
  \centering

\caption{\label{Tab-ResDust} Results of the fitting of dust models  } 
\begin{tabular}{rrrrrrrrr}
  \hline  \hline
  Identifier  &  $T_{\rm eff}$ &  Luminosity  & $\chi^2_{\rm r}$ &  $T_d$   & $\tau_{\rm v}$  & BIC & BIC &  BIC  \\ 
              &     (K)      &   (\lsol)     &                 &  (K)    &                 &     & &               \\ 
\hline
M31-PSO010.63 & 4625 & 33948 $\pm$ 1425 &  36.1 &  343 $\pm$  \phantom{0}47 & 0.793 $\pm$ 0.096  &  725 &  712 &  78137 \\ 
M31-GDR369249 & 4375 & 24688 $\pm$ 1516 &  41.3 &  609 $\pm$  148 & 0.495 $\pm$ 0.154            &  738 &  781 &   2529 \\ 
M31-VRJ004357 & 4000 & 28105 $\pm$ 2006 &  27.0 &  557 $\pm$  226 & 0.626 $\pm$ 0.244            &  453 &  430 &    964 \\ 
M31-PSO011.09 & 5500 & 23672 $\pm$ 1377 &  55.7 &  471 $\pm$  \phantom{0}60 & 0.760 $\pm$ 0.127  & 1296 & 1349 &  29370 \\ 
M31-VRJ004434 & 5000 & 37394 $\pm$ 1545 &  83.1 &  330 $\pm$  \phantom{0}45 & 1.037 $\pm$ 0.100  & 2188 & 2414 & 232000 \\ 
M33-013312    & 5125 & 29199 $\pm$ 5011 & 201.8 &  446 $\pm$  197 & 1.161 $\pm$ 0.412            & 3556 & 3805 &   6002  \\  
LMC-CEP-0619  & 4875 & 47329 $\pm$ 1315 &  54.8 &  925 $\pm$  111 & 0.082 $\pm$ 0.015            & 1730 & 1750 &  29766 \\ 
II Car        & 3600 & \phantom{0}9312 $\pm$ \phantom{1}592 & 164.3 & 1858 $\pm$  502 & 0.527 $\pm$ 0.218 & 3014 & 4077 & 5276 \\ 
 \hline
\end{tabular} 
\tablefoot{
Column~1. The identifier used in this paper.
Column~2. (photometric) effective temperature with error bar from the SED fitting.
Column~3. Luminosity with internal error bar from the SED fitting (for the fixed distance).
Column~4. Reduced chi-squared of the fit.
Column~5. Dust temperature at the inner radius.
Column~6. Dust optical depth in the $V$ band. The errors in $T_d$ and $\tau_{\rm v}$ are scaled to   $\chi^2_{\rm r}=1$.
  Column~7. BIC of this model.
Column~8. BIC of the model without dust and removing outliers (The model in Tab.~\ref{Tab-Res}).
Column~9. BIC of a model without dust and including all data points. 
}
\end{table*}

\subsection{Alternative models for the MW}
\label{SS:ALT}

For objects in the Bulge and MW, uncertainties in the adopted distance and reddening are important limitations
in deriving accurate luminosities and photometric effective temperatures (also see G20).
Table~\ref{Tab-Res} includes the distance and $A_{\rm V}$ estimates from \citet{Anders22} based on the StarHorse code \citep{Queiroz18}
that are independent from the initially adopted distances from \citet{BJ21} and the 3D reddening model.

\begin{figure}
  \centering

\begin{minipage}{0.49\textwidth}
\resizebox{\hsize}{!}{\includegraphics{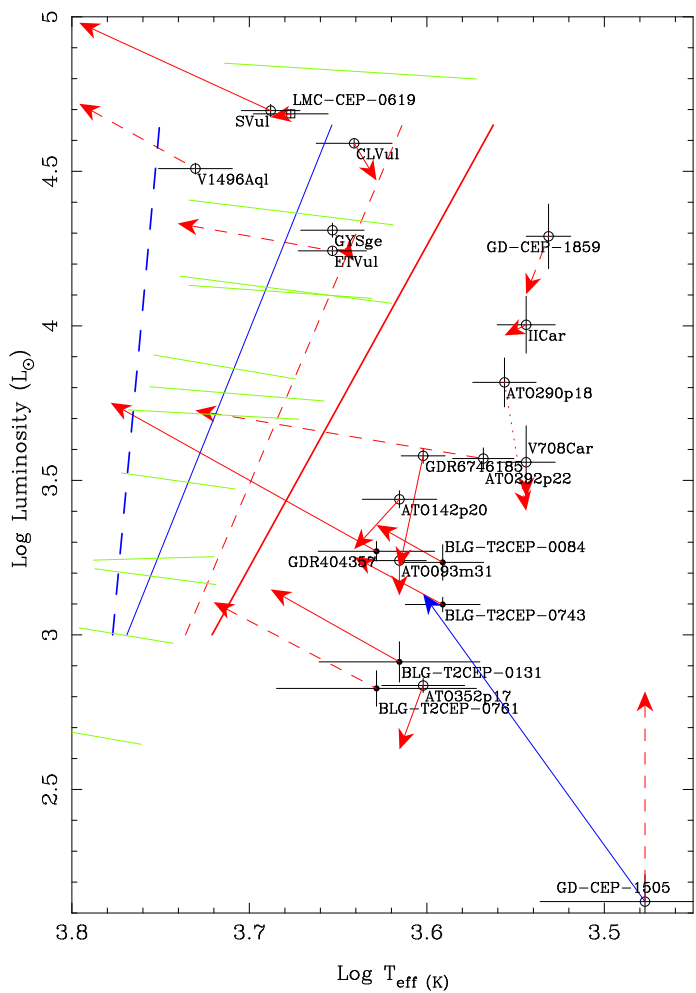}}
\end{minipage}

\caption{Hertzsprung--Russell diagram.
The symbols and lines largely follow Fig.~\ref{Fig:HRD}.
The standard models are connected to the best-fitting alternative models (i.e. with dust or alternative distances and reddenings)
by a red line with arrow. The arrow is
dashed        when the alternative model has a reduced $\chi^2$ that is lower by 10\% or more than that of the standard model,
dot-dashed when the alternative model has a reduced $\chi^2$ that is larger by 10\% or more, and solid otherwise.
The blue line for GD-CEP-1505 indicates yet another alternative model (see text).
}
\label{Fig:HRDALT}
\end{figure}

Models for the 20 MW and BUL stars were rerun based on the parameters from StarHorse or, when not available, on plausible
estimates based on general 1$\sigma$ error bars in the distance and on plausible extrapolations of the reddening.
The finally adopted parameters and the resulting best fits are listed in Table~\ref{Tab:ResAlt}.
Figure~\ref{Fig:HRDALT} compares the standard models with the best-fitting models including dust
or the alternative distances and reddenings.
Especially for some of the MW objects the change in luminosity and effective temperature are large (e.g. for GDR404357),
while the fit quality remains almost unchanged, indicating that the two parameters are degenerate.
While for some stars the alternative models move the object closer to or inside the IS, the location in the HRD of S Vul is moved outside the IS.
A special case is GD-CEP-1505. The alternative model is an improvement, but the fit is still the poorest
of all stars. Fixing the effective temperature to 4000~K and choosing a distance of 4.0~kpc will result
in a luminosity of 1350~\lsol, a position consistent with the other T2Cs in the sample and the luminosity predicted
for its period, but requires a very high reddening of $A_{\rm V} = 11$ for its Galactic position of $l= +47.09$, $b= +0.90$.
  Table~\ref{Tab:ResAlt} also includes some relevant quantities from the \G\ main catalogue, namely the $G$-band magnitude, the parallax, the
  goodness-of-fit parameter (GoF, expected to follow a Gaussian distribution centred on zero and with width unity), 
  and the renormalised unit weight error (expected to be centred on unity).
  
  The standard adopted distance from \citet{BJ21} uses the observed parallax, the parallax zero point correction from \citet{GEDR3_LindegrenZP},
  and a prior constructed from a three-dimensional model of the Galaxy to determine the distance and error.
  A few CCs are located outside the IS, which could point to a distance that is different from that in  \citet{BJ21} or the alternative distance.
  The parallax zero point correction is more uncertain for $G$ magnitudes $\less$12.5 (e.g. \citealt{CruzReyes23}), the significance of the
  parallaxes, $\pi/\sigma_{\pi}$, is often
  only 1-3 so that the prior will have a large impact on the derived distance, and some of the astrometric solutions are poor (RUWE $\more$ 1.4 or $\mid$ GoF $\mid \more 3$).
  \G\ DR4 is expected to deliver more accurate parallaxes that could resolve these issues.

\subsection{Period-luminosity and period-radius relations}
\label{SS:PL}

\begin{figure}
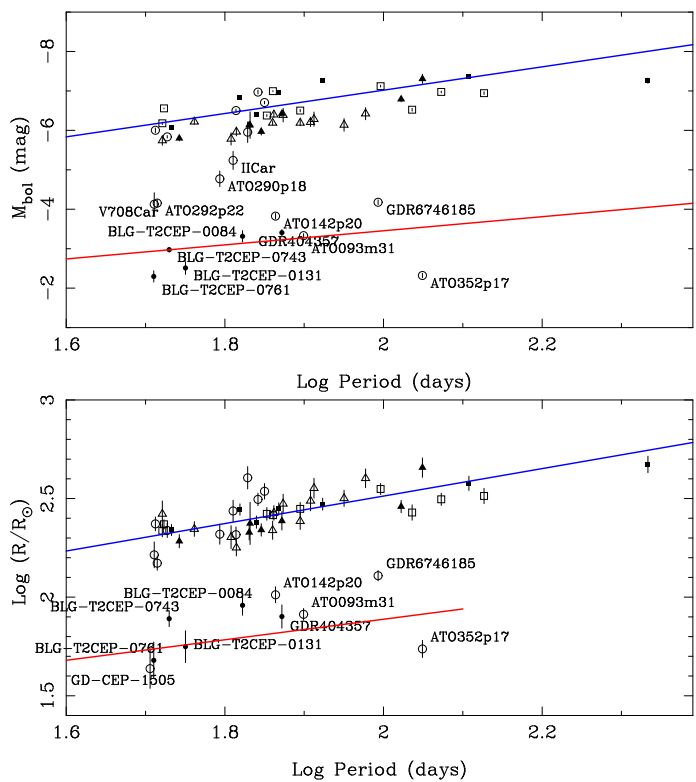

  \centering

\begin{minipage}{0.49\textwidth}
\resizebox{\hsize}{!}{\includegraphics{PL_Mbol_LogP.ps}}
\resizebox{\hsize}{!}{\includegraphics{Rad_Per.ps}}
\end{minipage}

\caption{Period-$M_{\rm bol}$ and $PR$ relations.
The error bars in $M_{\rm bol}$ are plotted but are typically smaller than the symbol size.
The symbols are follows: filled squares (SMC), open squares (LMC), open triangles (M31),
filled triangles (M33), filled circles (Galactic Bulge), and open circles (MW).
Stars located outside the bulk of objects are identified.
The blue lines give the relation for LMC CCs from \cite{GrLub}
while the red lines give the recommended solution for LMC T2Cs from \citet{GrJu17b}.
GD-CEP-1505 is located outside the upper plot at $\log P \sim 1.7$ and $M_{\rm bol} \sim -0.6$~mag.
}
\label{Fig:PL}
\end{figure}

Figure~\ref{Fig:PL} show the $PL$ and $PR$ relations based on the standard model results, together
with the $PL$ and $PR$ relations for CCs and T2Cs from \citet{GrLub} and \citet{GrJu17b} for the LMC.
The alternative models have not been plotted as they do not change the overall picture.
These relations confirm largely what is also seen in the HRD.
Some stars are located in these diagrams in positions consistent with their being T2Cs, but for some
this is true for the $PR$ diagram, but not for the $PL$ diagram.
There is more scatter in the relations for T2C than for CCs, but this is related to the fact that most
of the CCs are located in external galaxies with better defined distances and reddenings
than for the MW and Bulge objects.

Table~\ref{Tab:T2C} shows the classification of the 20 MW and Bulge objects based on the position in the HRD, the $PL$, and the $PR$ diagrams.
All ten that were previously classified or re-classified by the \G\ team as T2C are confirmed as such.
GD-CEP-1505 is most definitely not a CC. It is most likely a T2C, but for its effective temperature and luminosity to be consistent with this
requires a distance and reddening that is very different from that derived in the literature.
Six stars are certainly CCs, and for four stars the results are not conclusive.

\begin{table}
  \centering

\setlength{\tabcolsep}{1.6mm}
\caption{T2C and CC classification}
\begin{tabular}{rlrrrrcr} \hline \hline 
Name/Identifier & Type        & Type       &   Type            & Type \\
              &  (literature) & ($PR$)     &   ($PL$) & (HRD)   \\
  \hline   

BLG-T2CEP-0743 & DCEP * &  T2C &  T2C  & T2C   \\ 
BLG-T2CEP-0084 & DCEP * &  T2C &  T2C  & T2C  \\ 
BLG-T2CEP-0761 & DCEP * &  T2C &  T2C & T2C   \\ 
BLG-T2CEP-0131 & DCEP * &  T2C &  T2C  & T2C  \\ 
ATO093m31   &  DCEP * &  T2C  &  T2C  & T2C  \\ 
ATO142p20   &  DCEP * &  T2C &  T2C  & T2C  \\ 
GDR404357   &  DCEP * & T2C &  T2C  & T2C  \\ 
GDR6746185  &  DCEP * & T2C &  T2C  & T2C  \\ 
ATO352p17   &  DCEP * &  T2C &  T2C   & T2C  \\ 
GD-CEP-1505 &  T2CEP &   T2C &  non-CC  & non-CC  \\ 
V708 Car    &  DCEP   & CC &  T2C  & T2C \\ 
ATO292p22   &  DCEP  & CC &  T2C  & T2C  \\ 
II Car      &  DCEP   & CC & ?  & T2C \\ 
ATO290p18   &  DCEP  & CC & ? & T2C  \\ 
GD-CEP-1859 &  DCEP  & CC & CC  & T2C \\ 
V1496Aql    &  DCEP  & CC & CC & CC  \\ 
GY Sge      &  DCEP  & CC & CC & CC   \\ 
ET Vul      &  DCEP  & CC & CC & CC  \\ 
CL Vul      &  DCEP  & CC & CC & CC \\ 
S Vul       &  DCEP  &  CC & CC & CC \\ 
\hline
\end{tabular} 
\tablefoot{
  Copied from Table~\ref{Tab:Sample}.
  Column 2 lists the class from the GDR3 {\tt vari\_cepheid} table.
  An asterisk indicates that the star was re-classified as a T2C (see Table~6 in \citealt{RipepiDR3Cep22}).
  Columns 3, 4 and 5 give the type based on the location in the $PR$, $PL$ and HRD diagrams as derived in the present paper.
}
\label{Tab:T2C}
\end{table}

\section{Discussion and summary}
\label{S:Dis}

Table~\ref{Tab:Verd} summarises the results of the SED fitting of G20, \citet{GrLub}, and this paper in terms of
the likely presence of IR excess emission based on the SED fitting.
The numbers for the MW for $P>$ 50~d depend on whether II Car and some of the other objects are considered
CC or T2C (see Table~\ref{Tab:T2C}).

The fraction of CCs with IR excess appears small with the notable exception of the shorter period MW objects ($\sim$5\%).
In view of the Hubble tension (e.g. \citealt{HT2024}) it is an interesting question whether the presence of IR excess
could impact the derivation of the CC PL relation, especially if the effect were different in the
calibrating galaxies and in the galaxies the relation is applied to.
The current analysis suggests that the impact should be low at best.
Among the Galactic sample studied by \citet{RiessGEDR3} in their preferred {\sc HST} F555W, F814W, and F160W filter
system to calibrate the PL-relation, only S TrA and HW Car possibly have an IR excess, and
LMC619 is not among the  calibrating sample of 70 LMC CCs studied in \citet{Riess19}. 

Nevertheless, the presence and the origin of an IR excess remains intriguing and requires further study.
Although it was assumed in the modelling that this is due to dust this origin is problematic (how 
1000~K dust can form around a 6000~K central star; see discussion in G20) and 
free-free emission from ionised gas seems a viable alternative \citet{Hocde20,Hocde20b,HocdeKam25}.
Interferometric observations of the MW stars that are claimed to have IR excess emission based on SED modelling
would provide additional constraints.
An alternative would be to obtain MIR spectroscopy which would also be done for CCs in the LMC.
Even at a low resolution of $\sim$50 any silicate dust feature would become detectable and, if absent, any continuum emission
over that expected from the stellar photosphere would put constraints on the underlying mechanism.

\begin{table}
  \centering

\caption{Detection of IR excess}
\begin{tabular}{cccccrrrr} \hline \hline 
  Period & SMC & LMC & MW & M31+M33 \\
  (d)    &   & & &  \\
  \hline   
  $>50$ &  0/7  & 1/9   & 0/5-1/8 & 0/19 \\
  $<50$ &  0/72 & 0/134 & 16/350  & n.a. \\
\hline
\end{tabular} 
\label{Tab:Verd}
\end{table}

\section{Data availability}
The complete set of SEDs of the standard models and the alternative models is available at http://doi.org/10.5281/zenodo.15422721.

\begin{acknowledgements}
This work has made use of data from the European Space Agency (ESA) mission {\it Gaia} 
(\url{http://www.cosmos.esa.int/gaia}), processed by the {\it Gaia} Data Processing and Analysis Consortium 
(DPAC, \url{http://www.cosmos.esa.int/web/gaia/dpac/consortium}). 
Funding for the DPAC has been provided by national institutions, in particular
the institutions participating in the {\it Gaia} Multilateral Agreement.
This research has used data, tools or materials developed as part of the EXPLORE project that has received funding from the
European Union’s Horizon 2020 research and innovation programme under grant agreement No 101004214.
This research uses services or data provided by the Astro Data Lab at NSF's National Optical-Infrared Astronomy Research Laboratory.
NOIRLab is operated by the Association of Universities for Research in Astronomy (AURA), Inc. under a cooperative agreement with the National Science Foundation.
This research has made use of the SIMBAD database and the VizieR catalogue access tool 
operated at CDS, Strasbourg, France.
\end{acknowledgements}

\bibliographystyle{aa}
\bibliography{references.bib}

\clearpage
\onecolumn
\begin{appendix}
\section{Sample and fitting results}



\begin{table}[h]
  \centering

\footnotesize
\setlength{\tabcolsep}{1.9mm}
\caption{Sample of stars}
\begin{tabular}{rccrlrrrrl} \hline \hline 
Name/Identifier & RA       & Dec        & Period               & Type &  Period & Type & $L_{\rm CC}$ &   $L_{\rm T2C}$ & Remark \\
              &  (deg)   &  (deg)     & (d)                    &      & (d)     &      &  (\lsol)    &  (\lsol)    &  \\
  \hline   

M31-PSO009.76 & \phantom{0}9.765076 & 40.511659  &  64.236 $\pm$  0.079 &  DCEP  &   &  & 30790 &  1473 \\ 
M31-PSO010.11 & 10.119902 & 41.077721  &  52.622 $\pm$  0.046 &  DCEP  &   &  & 24314 &  1269 \\ 
M31-PSO010.19 & 10.198750 & 41.032396  &  57.743 $\pm$  0.052 &  DCEP  &   &  & 27140 &  1360 \\ 
M31-LGGS00415 & 10.313919 & 40.840238  &  94.933 $\pm$  0.148 &  DCEP  &   &  & 48894 &  1973 \\ 
M31-GDR381279 & 10.474412 & 41.451470  &  89.184 $\pm$  0.316 &  DCEP  &   &  & 45409 &  1883 \\ 
M31-PSO010.63 & 10.632660 & 41.486208  &  74.719 $\pm$  0.084 &  DCEP  &   &  & 36825 &  1649 \\ 
M31-GDR369249 & 10.768002 & 41.076318  &  80.914 $\pm$  0.118 &  DCEP  &   &  & 40467 &  1751 \\ 
M31-VRJ004357 & 10.991504 & 41.769738  &  81.734 $\pm$  0.207 &  DCEP  &   &  & 40953 &  1764 \\ 
M31-PSO011.09 & 11.091752 & 41.922604  &  65.243 $\pm$  0.024 &  DCEP  &   &  & 31362 &  1490 \\ 
M31-VRJ004434 & 11.143209 & 41.883533  &  72.749 $\pm$  0.090 &  DCEP  &   &  & 35678 &  1617 \\ 
M31-GDR387319 & 11.285518 & 42.099396  &  78.513 $\pm$  0.084 &  DCEP  &   &  & 39049 &  1712 \\ 
M31-GDR375422 & 11.889116 & 42.404674  &  72.481 $\pm$  0.036 &  DCEP  &   &  & 35522 &  1612 \\ 
SMC-CEP-0417 & 10.431045 & -73.723307  & 128.134 $\pm$  0.107 &  DCEP  & 128.197 $\pm$  0.303 &  F  & 69739 &  2469 & HV 821 \\ 
SMC-CEP-0921 & 11.721307 & -72.714377  &  65.925 $\pm$  0.042 &  DCEP  &  65.937 $\pm$  0.010 &  F  & 31751 &  1502 \\ 
SMC-CEP-1502 & 12.619999 & -72.752564  &  83.811 $\pm$  0.025 &  DCEP  &  84.300 $\pm$  0.046 &  F  & 42188 &  1797 & HV 829 \\ 
SMC-CEP-1977 & 13.236712 & -71.917574  &  69.157 $\pm$  0.039 &  DCEP  &  68.987 $\pm$  0.013 &  F  & 33602 &  1557 \\ 
SMC-CEP-2099 & 13.427702 & -72.287126  &  73.794 $\pm$  0.044 &  DCEP  &  73.621 $\pm$  0.036 &  F  & 36286 &  1634 \\ 
SMC-CEP-3611 & 16.064553 & -72.755640  & 215.479 $\pm$  1.781 &  DCEP  & 208.799 $\pm$  0.348 &  F  & 129048 &  3642 & HV 1956 \\ 
SMC-Dachs3-5 & 18.325909 & -73.363507  &  54.036 $\pm$  0.013 &  DCEP  &   &  & 25090 &  1294 \\ 
M33-013253 & 23.224087 & 30.590259  & 111.950 $\pm$  0.512 &  DCEP  &   &  & 59435 &  2232 \\ 
M33-013305 & 23.273916 & 30.622201  & 105.246 $\pm$  0.122 &  DCEP  &   &  & 55245 &  2131 \\ 
M33-013312 & 23.301018 & 30.646657  &  67.893 $\pm$  0.119 &  DCEP  &   &  & 32876 &  1535 \\ 
M33-013331 & 23.379284 & 30.528734  &  55.275 $\pm$  0.081 &  DCEP  &   &  & 25772 &  1316 \\ 
M33-013343 & 23.432750 & 30.545848  &  74.444 $\pm$  0.080 &  DCEP  &   &  & 36664 &  1645 \\ 
M33-V00021 & 23.465930 & 30.664079  &  67.711 $\pm$  0.284 &  DCEP  &   &  & 32772 &  1532 \\ 
M33-013405 & 23.521359 & 30.647512  &  70.130 $\pm$  0.080 &  DCEP  &   &  & 34162 &  1573 \\ 
LMC-CEP-4628 & 73.918991 & -66.428700  &  99.156 $\pm$  0.063 &  DCEP  &  99.200 $\pm$  0.500 &  F  & 51480 &  2038 & HV 5497 \\ 
LMC-CEP-4629 & 74.111650 & -64.694517  & 108.654 $\pm$  0.159 &  DCEP  & 108.700 $\pm$  0.500 &  F  & 57369 &  2183 & HV 2883 \\ 
LMC-CEP-0619 & 75.031541 & -68.450018  & 133.879 $\pm$  0.342 &  DCEP  & 133.779 $\pm$  0.152 &  F  & 73456 &  2551 & HV 883 \\ 
LMC-CEP-0992 & 76.816636 & -68.883485  &  52.875 $\pm$  0.032 &  DCEP  &  52.874 $\pm$  0.532 &  F  & 24453 &  1273 \\ 
LMC-CEP-1591 & 79.877126 & -68.686032  & 118.239 $\pm$  0.372 &  DCEP  & 118.624 $\pm$  0.266 &  F  & 63409 &  2325 & HV 2447 \\ 
LMC-CEP-2253 & 82.840800 & -70.957083  &  52.624 $\pm$  0.026 &  DCEP  &  52.374 $\pm$  0.034 &  F  & 24315 &  1269 \\ 
LMC-Dachs2-24 & 84.004780 & -66.844926  &  71.272 $\pm$  0.025 &  DCEP  &   &  & 34822 &  1592 \\ 
LMC-CEP-4689 & 85.947133 & -66.585814  &   &  &  78.508 $\pm$  0.377 &  F  & 39046 &  1712 \\ 
LMC-CEP-4691 & 86.210752 & -67.494585  &  72.533 $\pm$  0.057 &  DCEP  &  73.897 $\pm$  0.164 &  F  & 35553 &  1613 \\
\multicolumn{9}{c}{Milky Way Cepheids} \\
ATO093m31 & 93.250231 & -31.306365  &  79.284 $\pm$  0.152 &  DCEP * &   &  & 39504 &  1724 \\ 
ATO142p20 & 142.429652 & 20.824279  &  73.081 $\pm$  0.175 &  DCEP * &   &  & 35871 &  1622 \\ 
V708 Car & 153.907807 & -59.551290  &  51.404 $\pm$  0.016 &  DCEP  &   &  & 23649 &  1247 \\ 
II Car & 162.204346 & -60.063040  &  64.624 $\pm$  0.031 &  DCEP  &   &  & 31010 &  1480 \\ 
GD-CEP-1859 & 206.308819 & -63.571024  &  67.418 $\pm$  0.063 &  DCEP  &  67.568 $\pm$  0.014 &  F  & 32604 &  1527 \\ 
BLG-T2CEP-0743 & 267.339493 & -32.422453  &  53.672 $\pm$  0.147 &  DCEP * &  52.957 $\pm$  0.009 &  RVTau  & 24890 &  1288 \\ 
BLG-T2CEP-0084 & 267.569466 & -30.504683  &  66.414 $\pm$  0.166 &  DCEP * &  66.431 $\pm$  0.010 &  RVTau  & 32030 &  1510 \\ 
BLG-T2CEP-0761 & 267.647358 & -32.545647  &  51.325 $\pm$  0.204 &  DCEP * &  51.281 $\pm$  0.009 &  RVTau  & 23606 &  1245 \\ 
BLG-T2CEP-0131 & 268.107085 & -29.449319  &  56.276 $\pm$  0.030 &  DCEP * &  56.964 $\pm$  0.006 &  RVTau  & 26326 &  1334 \\ 
GDR404357 & 269.111824 & -32.190456  &  74.463 $\pm$  0.067 &  DCEP * &   &  & 36676 &  1645 \\ 
V1496 Aql & 283.748048 & -0.076784  &  65.174 $\pm$  0.109 &  DCEP  &   &  & 31323 &  1489 \\ 
GD-CEP-1505 & 288.605198 & 12.992103  &  50.786 $\pm$  0.029 &  T2CEP &  50.604 $\pm$  0.026 &  F  & 23313 &  1236 \\ 
ATO290p18 & 290.918840 & 18.378956  &  62.168 $\pm$  0.018 &  DCEP  &   &  & 29620 &  1437 \\ 
GDR6746185 & 291.573786 & -31.521751  &  98.444 $\pm$  0.086 &  DCEP * &   &  & 51043 &  2027 \\ 
ATO292p22 & 292.079775 & 22.484192  &  51.844 $\pm$  0.042 &  DCEP  &   &  & 23889 &  1255 \\ 
GY Sge & 293.806771 & 19.202388  &  51.566 $\pm$  0.040 &  DCEP  &   &  & 23738 &  1250 \\ 
ET Vul & 293.833756 & 26.430225  &  53.375 $\pm$  0.022 &  DCEP  &   &  & 24727 &  1282 \\ 
CL Vul & 294.965439 & 22.269290  &  70.797 $\pm$  0.054 &  DCEP  &   &  & 34548 &  1584 \\ 
S Vul & 297.099176 & 27.286481  &  69.467 $\pm$  0.046 &  DCEP  &   &  & 33781 &  1562 \\ 
ATO352p17 & 352.919044 & 17.853825  & 111.980 $\pm$  0.595 &  DCEP * &   &  & 59454 &  2232 \\ 
\hline
\end{tabular} 
\tablefoot{
  Columns 4 and 5 give the period and class from  the GDR3 {\tt vari\_cepheid} table.
  An asterisk in Col.~5 indicates that the star was re-classified as a T2C (see Table~6 in \citealt{RipepiDR3Cep22}).
  Columns 6 and 7 give the period and class (F stands for fundamental mode pulsator) from OGLE.
  Columns 8 and 9 give the predicted luminosity for the period listed in column~4 (except for LMC-CEP-4689)
  and the PL-relation for CCs and T2Cs in the LMC from  \citet{GrLub} and \citet{GrJu17b}, respectively.
  The last column gives the Harvard variable number for the CCs in the MCs. 
}
\label{Tab:Sample}
\end{table}


\begin{table}
  \centering
  \small
  
  \caption{\label{Tab-Res} Results of the fitting without dust }  
\begin{tabular}{rrccrrcrrr}
  \hline  \hline
  Identifier &  $d$      &  $A_{\rm V}$    & $T_{\rm eff}$ &  Luminosity  & $\chi^2_{\rm r}$ &  SH$_d$ & SH$_{A_{\rm v}}$ & $d_{\rm max}$  \\ 
           &    (kpc)  &  (mag)          &     (K)      &   (\lsol)    &               &  (kpc)  & (mag)     &   (kpc)   \\ 
\hline
 M31-PSO009.76   & 761.18 $\pm$ (11.0) &  0.17 & 4625 $\pm$ 312 & 16668 $\pm$ 2581 & 58.4 &                       &                     &          \\ 
 M31-PSO010.11   & 754.41 $\pm$ (11.0) &  0.17 & 4000 $\pm$ 347 & 16056 $\pm$ 1866 & 47.2 &                       &                     &          \\ 
 M31-PSO010.19   & 757.09 $\pm$ (11.0) &  0.17 & 4875 $\pm$ 221 & 24892 $\pm$ 1554 & 19.7 &                       &                     &          \\ 
 M31-LGGS00415   & 764.20 $\pm$ (11.0) &  0.17 & 3800 $\pm$ 184 & 30045 $\pm$ 4577 & 33.5 &                       &                     &          \\ 
 M31-GDR381279   & 752.32 $\pm$ (11.0) &  0.17 & 4000 $\pm$ 134 & 23101 $\pm$ 3654 & 17.7 &                       &                     &          \\ 
 M31-PSO010.63   & 766.98 $\pm$ (11.0) &  0.17 & 4375 $\pm$ 198 & 29078 $\pm$ 5035 & 32.3 &                       &                     &          \\ 
 M31-GDR369249   & 767.51 $\pm$ (11.0) &  0.17 & 4125 $\pm$ 250 & 24643 $\pm$ 2638 & 39.1 &                       &                     &          \\ 
 M31-VRJ004357   & 754.68 $\pm$ (11.0) &  0.17 & 3900 $\pm$ 180 & 26440 $\pm$ 4805 & 23.5 &                       &                     &          \\ 
 M31-PSO011.09   & 752.91 $\pm$ (11.0) &  0.17 & 5125 $\pm$ 243 & 19667 $\pm$ 1995 & 59.2 &                       &                     &          \\ 
 M31-VRJ004434   & 754.86 $\pm$ (11.0) &  0.17 & 4625 $\pm$ 208 & 29360 $\pm$ 1937 & 84.7 &                       &                     &          \\ 
 M31-GDR387319   & 752.36 $\pm$ (11.0) &  0.17 & 4625 $\pm$ 237 & 24216 $\pm$ 1611 & 47.9 &                       &                     &          \\ 
 M31-GDR375422   & 756.56 $\pm$ (11.0) &  0.17 & 4875 $\pm$ 289 & 24191 $\pm$ 1252 & 63.2 &                       &                     &          \\ 
 SMC-CEP-0417    &  62.44 $\pm$ (0.94) &  0.08 & 4875 $\pm$ 221 & 72352 $\pm$ 2823 & 59.8 &                       &                     &          \\ 
 SMC-CEP-0921    &  62.44 $\pm$ (0.94) &  0.10 & 5000 $\pm$ 177 & 43383 $\pm$ 1032 & 40.2 &                       &                     &          \\ 
 SMC-CEP-1502    &  62.44 $\pm$ (0.94) &  0.12 & 5375 $\pm$ 198 & 65640 $\pm$ 2687 & 58.2 &                       &                     &          \\ 
 SMC-CEP-1977    &  62.44 $\pm$ (0.94) &  0.13 & 4875 $\pm$ 189 & 29258 $\pm$ \phantom{0}677 & 46.5 &                       &                     &          \\ 
 SMC-CEP-2099    &  62.44 $\pm$ (0.94) &  0.12 & 5125 $\pm$ 189 & 48818 $\pm$ 1068 & 37.5 &                       &                     &          \\ 
 SMC-CEP-3611    &  62.44 $\pm$ (0.94) &  0.10 & 4250 $\pm$ 221 & 64918 $\pm$ 4074 & 128.8 &                       &                     &          \\ 
 SMC-Dachs3-5    &  62.44 $\pm$ (0.94) &  0.23 & 4750 $\pm$ 168 & 21979 $\pm$ \phantom{0}365 & 30.0 &                       &                     &          \\ 
 M33-013253      & 844.54 $\pm$ (11.0) &  0.17 & 4375 $\pm$ 253 & 67771 $\pm$ 7606 & 95.2 &                       &                     &          \\ 
 M33-013305      & 843.49 $\pm$ (11.0) &  0.17 & 4875 $\pm$ 177 & 42087 $\pm$ 2642 & 31.3 &                       &                     &          \\ 
 M33-013312      & 842.95 $\pm$ (11.0) &  0.17 & 4625 $\pm$ 627 & 22999 $\pm$ 8364 & 193.7 &                       &                     &          \\ 
 M33-013331      & 843.10 $\pm$ (11.0) &  0.17 & 4750 $\pm$ 189 & 16893 $\pm$ 1049 & 45.5 &                       &                     &          \\ 
 M33-013343      & 842.40 $\pm$ (11.0) &  0.17 & 4875 $\pm$ 289 & 30123 $\pm$ 2067 & 100.5 &                       &                     &          \\ 
 M33-V00021      & 840.00 $\pm$ (11.0) &  0.17 & 4875 $\pm$ 226 & 22999 $\pm$ 1364 & 46.2 &                       &                     &          \\ 
 M33-013405      & 841.16 $\pm$ (11.0) &  0.17 & 4625 $\pm$ 153 & 19762 $\pm$ 1690 & 63.0 &                       &                     &          \\ 
 LMC-CEP-4628    &  49.53 $\pm$ (0.55) &  0.31 & 4750 $\pm$ 168 & 57131 $\pm$ 1199 & 24.7 &                       &                     &          \\ 
 LMC-CEP-4629    &  49.14 $\pm$ (0.55) &  0.15 & 4750 $\pm$ 237 & 32987 $\pm$ 1120 & 69.6 &                       &                     &          \\ 
 LMC-CEP-0619    &  49.89 $\pm$ (0.55) &  0.28 & 4750 $\pm$ 236 & 48544 $\pm$ 1538 & 51.3 &                       &                     &          \\ 
 LMC-CEP-0992    &  49.80 $\pm$ (0.55) &  0.22 & 5125 $\pm$ 144 & 33980 $\pm$ \phantom{0}366 & 9.5 &                       &                     &          \\ 
 LMC-CEP-1591    &  49.43 $\pm$ (0.55) &  0.30 & 4875 $\pm$ 189 & 49882 $\pm$ 1085 & 32.9 &                       &                     &          \\ 
 LMC-CEP-2253    &  49.88 $\pm$ (0.55) &  0.31 & 4875 $\pm$ 189 & 23982 $\pm$ \phantom{0}689 & 23.3 &                       &                     &          \\ 
 LMC-Dachs2-24   &  48.53 $\pm$ (0.55) &  0.23 & 4625 $\pm$ 189 & 28762 $\pm$ \phantom{0}726 & 17.8 &                       &                     &          \\ 
 LMC-CEP-4689    &  48.31 $\pm$ (0.55) &  0.21 & 4625 $\pm$ 188 & 32283 $\pm$ \phantom{0}996 & 48.4 &                       &                     &          \\ 
 LMC-CEP-4691    &  48.53 $\pm$ (0.55) &  0.33 & 5375 $\pm$ 125 & 50872 $\pm$ \phantom{0}599 & 8.6 &                       &                     &          \\ 
 \multicolumn{9}{c}{Milky Way Cepheids} \\
ATO093m31       &  15.78 $\pm$ (2.58) &  0.10 & 4125 $\pm$ 145 &  1740 $\pm$ \phantom{00}87 & 31.2 &                         &                     & 1.12 \\ 
 ATO142p20       &   9.72 $\pm$ (2.43) &  0.08 & 4125 $\pm$ 204 &  2747 $\pm$ \phantom{0}181 & 55.2 &     7.36 $\pm$     0.15 &   0.51 $\pm$   0.01 & 0.59 \\ 
 V708 Car         &   3.87 $\pm$ (0.33) &  1.33 & 3500 $\pm$ 134 &  3619 $\pm$ 1130 & 255.0 &                         &                     & 3.87 \\ 
 II Car          &   7.66 $\pm$ (1.24) &  1.41 & 3500 $\pm$ 134 & 10082 $\pm$ 2377 & 199.4 &     5.11 $\pm$     1.75 &   2.57 $\pm$   0.42 & 5.27 \\ 
 GD-CEP-1859     &  10.62 $\pm$ (3.02) &  3.21 & 3400 $\pm$ 100 & 19488 $\pm$ 5306 & 230.2 &                         &                     & 6.43 \\ 
 BLG-T2CEP-0743  &   8.28 $\pm$ (0.03) &  4.10 & 3900 $\pm$ 194 &  1254 $\pm$ \phantom{00}70 & 83.5 &                         &                     & 5.01 \\ 
 BLG-T2CEP-0084  &   8.28 $\pm$ (0.03) &  4.28 & 3900 $\pm$ 212 &  1716 $\pm$ \phantom{0}243 & 139.1 &    10.99 $\pm$     0.98 &   6.31 $\pm$   0.06 & 5.01 \\ 
 BLG-T2CEP-0761  &   8.28 $\pm$ (0.03) &  3.74 & 4250 $\pm$ 588 &   671 $\pm$ \phantom{00}95 & 478.5 &     7.99 $\pm$     1.54 &   4.82 $\pm$   2.12 & 5.02 \\ 
 BLG-T2CEP-0131  &   8.28 $\pm$ (0.03) &  3.85 & 4125 $\pm$ 453 &   818 $\pm$ \phantom{0}133 & 486.7 &                         &                     & 5.01 \\ 
 GDR404357       &   8.28 $\pm$ (0.03) &  2.96 & 4250 $\pm$ 333 &  1865 $\pm$ \phantom{0}142 & 132.5 &                         &                     & 5.02 \\ 
 V1496 Aql       &   3.54 $\pm$ (0.32) &  3.95 & 5375 $\pm$ 264 & 32271 $\pm$ 1309 & 69.8 &     3.26 $\pm$     0.18 &   4.85 $\pm$   0.35 & 3.54 \\ 
 GD-CEP-1505     &   4.05 $\pm$ (2.21) &  3.59 & 3000 $\pm$ 438 &   137 $\pm$ \phantom{00}30 & 1940.0 &                         &                     & 4.05 \\ 
 ATO290p18       &   7.82 $\pm$ (2.64) &  5.44 & 3600 $\pm$ 151 &  6562 $\pm$ 1315 & 204.9 &                         &                     & 6.28 \\ 
 GDR6746185      &   9.16 $\pm$ (1.56) &  0.32 & 4000 $\pm$ 115 &  3795 $\pm$ \phantom{0}218 & 22.5 &     5.70 $\pm$     0.06 &   0.58 $\pm$   0.01 & 1.16 \\ 
 ATO292p22       &  12.55 $\pm$ (2.18) &  3.01 & 3700 $\pm$ 150 &  3720 $\pm$ \phantom{0}307 & 60.7 &     8.51 $\pm$     0.91 &   5.59 $\pm$   0.50 & 5.97 \\ 
 GY Sge          &   2.89 $\pm$ (0.18) &  3.51 & 4500 $\pm$ 189 & 20396 $\pm$ 1113 & 43.7 &                         &                     & 2.89 \\ 
 ET Vul          &  13.94 $\pm$ (2.52) &  1.93 & 4500 $\pm$ 204 & 17489 $\pm$ \phantom{0}544 & 53.7 &                         &                     & 5.72 \\ 
 CL Vul          &   4.28 $\pm$ (0.38) &  4.61 & 4375 $\pm$ 221 & 38958 $\pm$ 1242 & 43.4 &                         &                     & 4.28 \\ 
 S Vul           &   4.32 $\pm$ (0.33) &  2.58 & 4875 $\pm$ 189 & 49736 $\pm$ 2351 & 36.5 &     3.99 $\pm$     0.44 &   3.79 $\pm$   1.69 & 4.32 \\ 
 ATO352p17       &  11.87 $\pm$ (2.65) &  0.20 & 4000 $\pm$ 221 &   686 $\pm$  \phantom{00}37 & 46.3 &                         &                     & 0.62 \\ 
 \hline
\end{tabular} 
\tablefoot{
Column~1. The identifier used in this paper.
Column~2. The adopted distance.
The error in the distance is listed between parenthesis to indicate that this error is not included in the luminosity error estimate.
Column~3. The adopted reddening value $A_{\rm V}$.
Column~4. (photometric) effective temperature with error bar from the SED fitting.
Column~5. Luminosity with error bar from the SED fitting (for the fixed distance).
Column~6. Reduced chi-squared of the fit.
Column~7. Distance with error bar from  StarHorse \citep{Anders22}.
Column~8. Reddening value $A_{\rm V}$ from  StarHorse. 
Column~9. The maximum distance to which the 3D reddening model is available.
}
\end{table}


\begin{table}
  \centering

\caption{Results of the fitting using alternative distances and reddenings, and some \G\ parameters}
\begin{tabular}{rrccrrr rrrr}
\hline \hline 
  Identifier &  $d$      &  $A_{\rm V}$    & $T_{\rm eff}$ &  Luminosity  & $\chi^2_{\rm r}$ & $G$  & $\pi \pm \sigma_{\pi}$ & GoF &  RUWE \\ 
             &    (kpc)  &  (mag)          &     (K)      &   (\lsol)    &                & (mag)  & (mas) & &  \\  \hline
 ATO093m31       &  13.20 &  0.30 & 4125 $\pm$ 188 &  1341 $\pm$   42 & 30.1 &  13.0 &   0.035 $\pm$   0.013 &  1.43 &  1.05 \\ 
 ATO142p20       &   7.40 &  0.50 & 4375 $\pm$ 204 &  1895 $\pm$   97 & 51.6 &  11.6 &   0.044 $\pm$   0.027 &  6.09 &  1.23 \\ 
 V708 Car         &   3.50 &  1.24 & 3500 $\pm$ 122 &  2791 $\pm$ 1315 & 282.4 &  10.7 &   0.226 $\pm$   0.023 &  1.08 &  1.04 \\ 
 II Car           &   5.10 &  2.60 & 3700 $\pm$ 225 &  6792 $\pm$  927 & 178.0 &  11.3 &   0.105 $\pm$   0.021 & -3.09 &  0.89 \\ 
 GD-CEP-1859     &   7.60 &  4.00 & 3500 $\pm$ 115 & 12649 $\pm$ 1769 & 154.0 &  12.5 &   0.070 $\pm$   0.038 & -1.91 &  0.94 \\ 
 BLG-T2CEP-0743  &   8.28 &  5.00 & 4375 $\pm$ 277 &  1786 $\pm$   70 & 82.0 &  15.1 &   0.064 $\pm$   0.031 & -2.21 &  0.91 \\ 
 BLG-T2CEP-0084  &   8.28 &  5.00 & 4250 $\pm$ 277 &  2268 $\pm$  198 & 139.3 &  14.9 &  -0.101 $\pm$   0.047 &  3.32 &  1.13 \\ 
 BLG-T2CEP-0761  &   8.28 &  5.00 & 5250 $\pm$ 356 &  1277 $\pm$  118 & 391.3 &  15.2 &   0.043 $\pm$   0.035 & -1.87 &  0.93 \\ 
 BLG-T2CEP-0131  &   8.28 &  5.00 & 4875 $\pm$ 315 &  1404 $\pm$  164 & 470.9 &  15.1 &   0.053 $\pm$   0.040 & -0.16 &  0.99 \\ 
 GDR404357       &   8.28 &  5.00 & 6000 $\pm$ 434 &  5638 $\pm$  481 & 136.5 &  13.5 &   0.026 $\pm$   0.024 & -1.18 &  0.95 \\ 
 V1496 Aql       &   3.30 &  4.90 & 6250 $\pm$ 335 & 52330 $\pm$ 2308 & 55.9 &   9.1 &   0.236 $\pm$   0.025 &  0.33 &  1.01 \\ 
 GD-CEP-1505     &   6.30 &  5.50 & 3000 $\pm$ 283 &   657 $\pm$  109 & 722.3 &  17.2 &   0.267 $\pm$   0.146 &  6.57 &  1.20 \\ 
  (idem)         & 4.0\phantom{0} & 11.0\phantom{0} & 4000\phantom{.} fixed & 1355 $\pm$ 54  & 79.0   & & &  \\
 ATO290p18       &   5.20 &  4.90 & 3500 $\pm$ 122 &  2520 $\pm$  619 & 255.9 &  14.1 &   0.049 $\pm$   0.052 &  2.43 &  1.10 \\ 
 GDR6746185      &   5.70 &  0.60 & 4125 $\pm$ 144 &  1654 $\pm$   50 & 21.0 &  11.2 &   0.079 $\pm$   0.024 &  5.66 &  1.30 \\ 
 ATO292p22       &   8.50 &  5.60 & 5375 $\pm$ 198 &  5323 $\pm$  193 & 46.5 &  14.0 &   0.016 $\pm$   0.016 & -2.60 &  0.92 \\ 
 GY Sge          &   2.70 &  3.30 & 4375 $\pm$ 189 & 16229 $\pm$  852 & 49.9 &   8.8 &   0.296 $\pm$   0.022 & -1.61 &  0.95 \\ 
 ET Vul          &  11.40 &  3.00 & 5500 $\pm$ 250 & 21437 $\pm$  827 & 40.0 &  11.3 &   0.040 $\pm$   0.017 &  3.35 &  1.10 \\ 
 CL Vul          &   3.90 &  4.40 & 4250 $\pm$ 204 & 29526 $\pm$  923 & 46.1 &   9.6 &   0.181 $\pm$   0.023 & -5.99 &  0.82 \\ 
 S Vul           &   4.00 &  3.80 & 6250 $\pm$ 312 & 95670 $\pm$ 7465 & 39.9 &   8.2 &   0.205 $\pm$   0.020 &  1.13 &  1.03 \\ 
 ATO352p17       &   9.20 &  0.30 & 4125 $\pm$ 204 &   426 $\pm$   \phantom{11}20 & 45.6 &  13.5 &   0.023 $\pm$   0.021 & 11.40 &  1.43 \\ 
 \hline
\end{tabular} 
\tablefoot{Columns~1-6 as in Table~\ref{Tab-Res}.
 Columns~7-10 parameters from the \G\ catalog; $G$-band magnitude, parallax with error, goodness-of-fit (GoF, {\tt astrometric\_gof\_al}),
  and the renormalised unit weight error ({\tt RUWE}).  
}
\label{Tab:ResAlt}
\end{table}

\FloatBarrier
\clearpage

\section{Sources of the photometry}

\begin{table}[h]
  
  \centering
  
  \caption{\label{Tab-Phot} Photometry used to construct the SEDs} 
    \begin{tabular}{llll}
  \hline  \hline
 Filters  & Instrument         &  Reference                 & Remark    \\ 
\hline
 UV    & GALEX               & \citet{Bianchi17} & \\
 $V,I$ & OGLE Shallow Survey &  \citet{Ulaczyk12,Ulaczyk13} & LMC \\
 $V,I$ & OGLE-{\sc iv} & \citet{Soszynski19,Soszynski17,Soszynski2020,Udalski2018} \\
 $B_{\rm p}$, $G$, $R_{\rm p}$ & {\it Gaia} & \citet{GaiaDR3Vallenari22, RipepiDR3Cep22} \\
 $U,B,V,R,I$ & & \citet{Berdnikov2008, Berdnikov2015} \\
 $U,B,V$ &  & \citet{SzabadosMitSU70,SzabadosCoKon76,SzabadosCoKon77,SzabadosCoKon96} \\
 $U,B,V$ &  & \citet{Madore1975,Eggen1977,MW1979} \\
 VBLUW   & Walraven & \citet{Pel76,GrLub} \\
 $u,g,r,i$ & VPHAS+ DR2 &  \citet{Drew14,VPHAS16} \\
 $u,g,r,i,z$ &  SMASH DR2 & \citet{Nidever21} & SMC \\
 $g,r,i,z$  &  Pan-STARRS DR1 &  \citet{PS1} \\
 $g,r$ & ZTF & \citet{Chen20} \\
 $B,V,g,r,i$ & APASS DR9 & \citet{HendenAPASS}\\
 $i$  & AST survey &  \citet{AST18}\\
 $U,B,V,R,I$ & & \citet{Massey16} & M31, M33 \\
 $B,V,I$ &  & \citet{Pellerin11} & M33 \\
 F555W, F814W, F160W & HST & \cite{Riess19, RiessGEDR3} \\
 \tablefootmark{a} & HST/PHAT & \citet{Williams14} & M31 \\
 \tablefootmark{a} & HST/PHATTER & \citet{Williams21} & M33 \\
 $Y,J,K$ & VMC  & \citet{Cioni11,Ripepi16,Ripepi22} \\
 $Z,Y,J,H,K$ & VVV DR5 &  \citet{Minniti2010} & \\
 $Y,J,H,K$ & VHS DR6 & \citet{McMahon13} \\
 $J,H,K$ & 2MASS, 2MASS-6X &   \citet{Cutri2003,Cutri12}\\
 $J,H,K$ &          & \citet{Laney1992, MP11} \\
 $I,J,K$ & DENIS &  \citet{DENIS05} \\
 $J,K$  &  & \citet{Neugent20} & M31 \\
 $J,H,K$ & & \citet{Javadi11} &  M33 \\
 S7, S11, L15\tablefootmark{b} & Akari & \citet{Ita2010, Kato12} &  \\ 
 9, 18~$\mu m$ & Akari & \cite{AKIRC10} \\
 W1, W2 & CatWISE & \citet{CatWISE21} \\
 W3, W4\tablefootmark{c} & AllWISE & \citet{Cutri_Allwise} \\
 3.6, 4.5~$\mu m$ & {\it Spitzer} IRAC & \citet{Chown21} \\
 5.8, 8.5~$\mu m$ & {\it Spitzer} IRAC & IPAC\tablefootmark{d} \\
 IRAC, MIPS 24~$\mu m$ &                  &  \citet{Khan15, Khan17} & M31, M33 \\
 IRAC, MIPS 24~$\mu m$ & GLIMPSE, MIPSGAL &  \citet{GLIMPSE09,MIPSGAL15} \\
 24~$\mu m$ & MIPS &                   IRSA\tablefootmark{e} & MCs \\
 A,C,D   & MSX & \citet{Egan03} & \\
 12~$\mu m$  & IRAS PSC & \citet{Beichmann85} \\ 
 70~$\mu m$  & {\it Herschel} PACS & \citet{HerschelPACS20} \\
 \hline
    \end{tabular}
\tablefoot{
\tablefoottext{a}{F275W, F336W, F475W, F814W, F110W, and F160W};
\tablefoottext{b}{For the S7, S11, and L15 filters only errors in the magnitudes were accepted of $<$0.15, $<$0.20, and $<$0.20~mag, respectively};
\tablefoottext{c}{In the W3 and W4 filters only errors in the magnitudes were accepted of $<$0.30, and $<$0.25~mag, respectively};
\tablefoottext{d}{VizieR catalog II/305/catalog};
\tablefoottext{e}{\url{https://irsa.ipac.caltech.edu/applications/Gator/}, the "SAGE MIPS 24 um Epoch 1 and Epoch 2 Full List"}.
}
\end{table}

\FloatBarrier
\clearpage

 \section{Additional figures}

\begin{figure}[h]
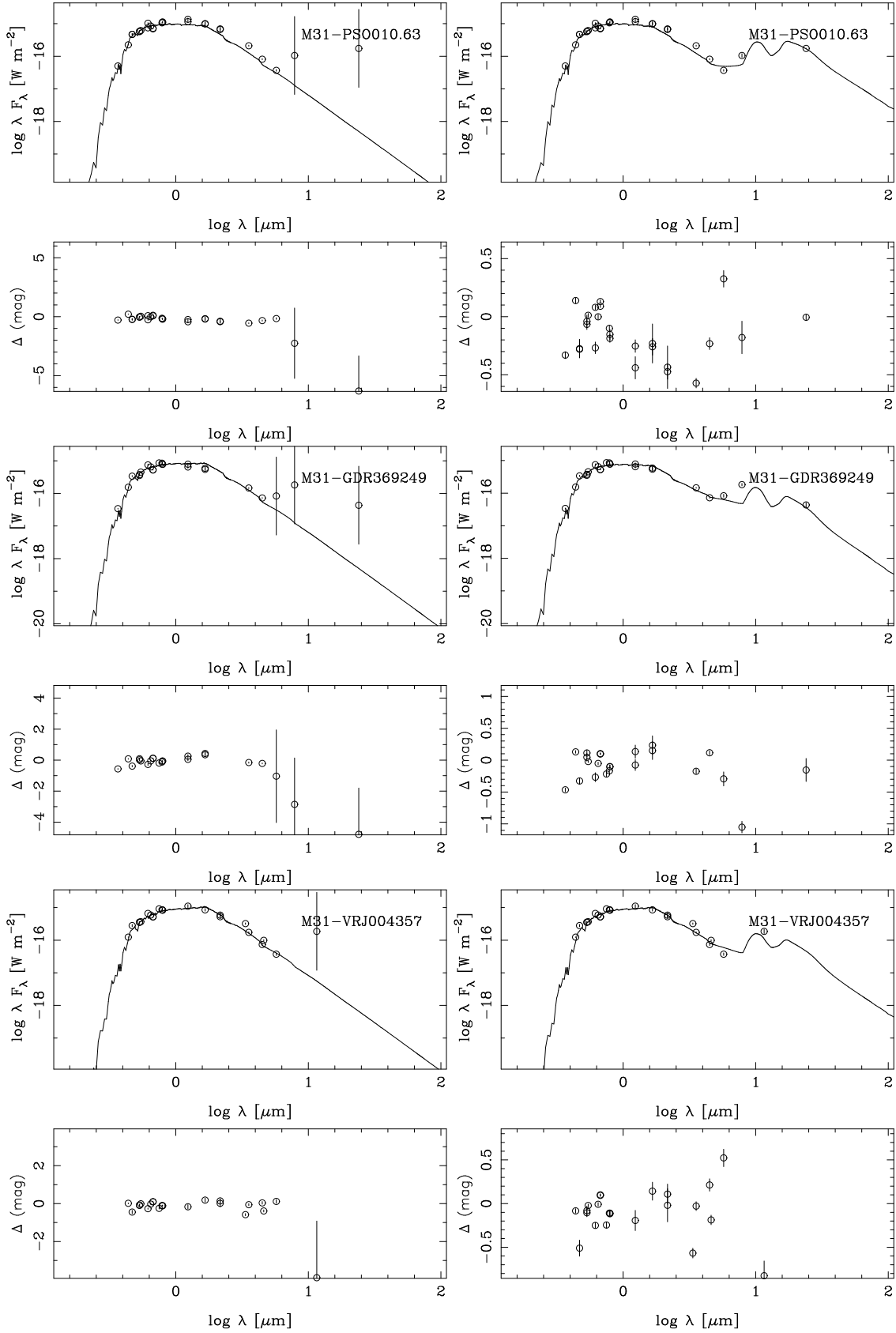


\begin{minipage}{0.40\textwidth}
\resizebox{\hsize}{!}{\includegraphics{M31-PSO010.63_sed.ps}}
\end{minipage}
\begin{minipage}{0.40\textwidth}
\resizebox{\hsize}{!}{\includegraphics{M31-PSO010.63_sed_dusty.ps}}
\end{minipage}

\begin{minipage}{0.40\textwidth}
\resizebox{\hsize}{!}{\includegraphics{M31-GDR369249_sed.ps}}
\end{minipage}
\begin{minipage}{0.40\textwidth}
\resizebox{\hsize}{!}{\includegraphics{M31-GDR369249_sed_dusty.ps}}
\end{minipage}

\begin{minipage}{0.40\textwidth}
\resizebox{\hsize}{!}{\includegraphics{M31-VRJ004357_sed.ps}}
\end{minipage}
\begin{minipage}{0.40\textwidth}
\resizebox{\hsize}{!}{\includegraphics{M31-VRJ004357_sed_dusty.ps}}
\end{minipage}

\caption{Best-fitting models assuming dust (right-hand) compared to no dust (left-hand panels) of the remaining 6 objects (cf.  Fig.~\ref{Fig:sed2}).
    Note the difference in the range of the ordinate in the left-hand and right-hand bottom panels.      
}
\label{Fig:seddusty}
\end{figure}

\setcounter{figure}{0}

\begin{figure}

\begin{minipage}{0.43\textwidth}
\resizebox{\hsize}{!}{\includegraphics{M31-PSO011.09_sed.ps}}
\end{minipage}
\begin{minipage}{0.43\textwidth}
\resizebox{\hsize}{!}{\includegraphics{M31-PSO011.09_sed_dusty.ps}}
\end{minipage}

\begin{minipage}{0.43\textwidth}
\resizebox{\hsize}{!}{\includegraphics{M33-013312_sed.ps}}
\end{minipage}
\begin{minipage}{0.43\textwidth}
\resizebox{\hsize}{!}{\includegraphics{M33-013312_sed_dusty.ps}}
\end{minipage}

\begin{minipage}{0.43\textwidth}
\resizebox{\hsize}{!}{\includegraphics{IICar_sed.ps}}
\end{minipage}
\begin{minipage}{0.43\textwidth}
\resizebox{\hsize}{!}{\includegraphics{IICar_sed_dusty.ps}}
\end{minipage}

\caption{continued.
}
\label{Fig:seddusty}
\end{figure}

  \begin{figure}[h]

\begin{minipage}{0.23\textwidth}
\resizebox{\hsize}{!}{\includegraphics{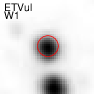}}
\end{minipage}
\begin{minipage}{0.23\textwidth}
\resizebox{\hsize}{!}{\includegraphics{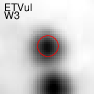}}
\end{minipage}
\begin{minipage}{0.02\textwidth}
\end{minipage}
\begin{minipage}{0.23\textwidth}
\resizebox{\hsize}{!}{\includegraphics{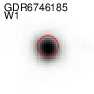}}
\end{minipage}
\begin{minipage}{0.23\textwidth}
\resizebox{\hsize}{!}{\includegraphics{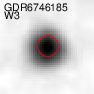}}
\end{minipage}

\begin{minipage}{0.23\textwidth}
\resizebox{\hsize}{!}{\includegraphics{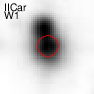}}
\end{minipage}
\begin{minipage}{0.23\textwidth}
\resizebox{\hsize}{!}{\includegraphics{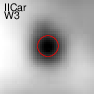}}
\end{minipage}
\begin{minipage}{0.02\textwidth}
  \phantom{.}
\end{minipage}
\begin{minipage}{0.23\textwidth}
\resizebox{\hsize}{!}{\includegraphics{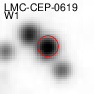}}
\end{minipage}
\begin{minipage}{0.23\textwidth}
\resizebox{\hsize}{!}{\includegraphics{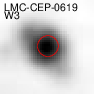}}
\end{minipage}

\begin{minipage}{0.23\textwidth}
\resizebox{\hsize}{!}{\includegraphics{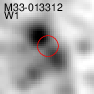}}
\end{minipage}
\begin{minipage}{0.23\textwidth}
\resizebox{\hsize}{!}{\includegraphics{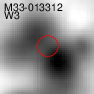}}
\end{minipage}
\begin{minipage}{0.02\textwidth}
  \phantom{.}
\end{minipage}
\begin{minipage}{0.23\textwidth}
\resizebox{\hsize}{!}{\includegraphics{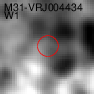}}
\end{minipage}
\begin{minipage}{0.23\textwidth}
\resizebox{\hsize}{!}{\includegraphics{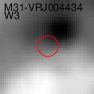}}
\end{minipage}

\begin{minipage}{0.23\textwidth}
\resizebox{\hsize}{!}{\includegraphics{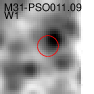}}
\end{minipage}
\begin{minipage}{0.23\textwidth}
\resizebox{\hsize}{!}{\includegraphics{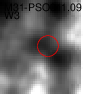}}
\end{minipage}
\begin{minipage}{0.02\textwidth}
  \phantom{.}
\end{minipage}
\begin{minipage}{0.23\textwidth}
\resizebox{\hsize}{!}{\includegraphics{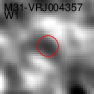}}
\end{minipage}
\begin{minipage}{0.23\textwidth}
\resizebox{\hsize}{!}{\includegraphics{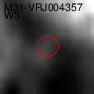}}
\end{minipage}

\begin{minipage}{0.23\textwidth}
\resizebox{\hsize}{!}{\includegraphics{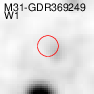}}
\end{minipage}
\begin{minipage}{0.23\textwidth}
\resizebox{\hsize}{!}{\includegraphics{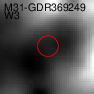}}
\end{minipage}
\begin{minipage}{0.02\textwidth}
  \phantom{.}
\end{minipage}
\begin{minipage}{0.23\textwidth}
\resizebox{\hsize}{!}{\includegraphics{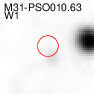}}
\end{minipage}
\begin{minipage}{0.23\textwidth}
\resizebox{\hsize}{!}{\includegraphics{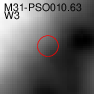}}
\end{minipage}

\caption{Cut-outs of about 1\arcmin x 1\arcmin\ (45x45 pixels of 1.37\arcsec) in the W1 and W3 filters
  centred on the CC. Cut levels are at the 0.5 and 99.5\% level.
  The red circle marks the nominal position and has a radius of 5 pixels, corresponding to approximately 1 FWHM of the point spread function.
  ET Vul and GDR 6746185 are plotted for comparison as MW stars not having an IR excess.
}
\label{Fig:WISE}
\end{figure}

\begin{figure}

\begin{minipage}{0.23\textwidth}
\resizebox{\hsize}{!}{\includegraphics{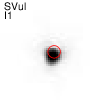}}
\end{minipage}
\begin{minipage}{0.23\textwidth}
\resizebox{\hsize}{!}{\includegraphics{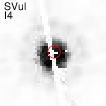}}
\end{minipage}
\begin{minipage}{0.02\textwidth}
\end{minipage}
\begin{minipage}{0.23\textwidth}
\resizebox{\hsize}{!}{\includegraphics{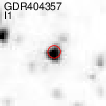}}
\end{minipage}
\begin{minipage}{0.23\textwidth}
\resizebox{\hsize}{!}{\includegraphics{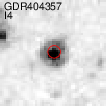}}
\end{minipage}

\begin{minipage}{0.23\textwidth}
\resizebox{\hsize}{!}{\includegraphics{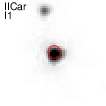}}
\end{minipage}
\begin{minipage}{0.23\textwidth}
\resizebox{\hsize}{!}{\includegraphics{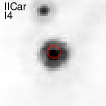}}
\end{minipage}
\begin{minipage}{0.02\textwidth}
  \phantom{.}
\end{minipage}
\begin{minipage}{0.23\textwidth}
\resizebox{\hsize}{!}{\includegraphics{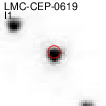}}
\end{minipage}
\begin{minipage}{0.23\textwidth}
\resizebox{\hsize}{!}{\includegraphics{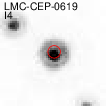}}
\end{minipage}

\begin{minipage}{0.23\textwidth}
\resizebox{\hsize}{!}{\includegraphics{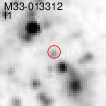}}
\end{minipage}
\begin{minipage}{0.23\textwidth}
\resizebox{\hsize}{!}{\includegraphics{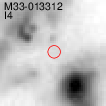}}
\end{minipage}
\begin{minipage}{0.02\textwidth}
  \phantom{.}
\end{minipage}
\begin{minipage}{0.23\textwidth}
\resizebox{\hsize}{!}{\includegraphics{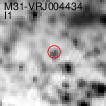}}
\end{minipage}
\begin{minipage}{0.23\textwidth}
\resizebox{\hsize}{!}{\includegraphics{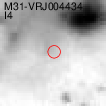}}
\end{minipage}

\begin{minipage}{0.23\textwidth}
\resizebox{\hsize}{!}{\includegraphics{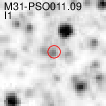}}
\end{minipage}
\begin{minipage}{0.23\textwidth}
\resizebox{\hsize}{!}{\includegraphics{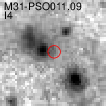}}
\end{minipage}
\begin{minipage}{0.02\textwidth}
  \phantom{.}
\end{minipage}
\begin{minipage}{0.23\textwidth}
\resizebox{\hsize}{!}{\includegraphics{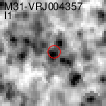}}
\end{minipage}
\begin{minipage}{0.23\textwidth}
\resizebox{\hsize}{!}{\includegraphics{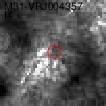}}
\end{minipage}

\begin{minipage}{0.23\textwidth}
\resizebox{\hsize}{!}{\includegraphics{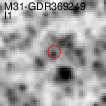}}
\end{minipage}
\begin{minipage}{0.23\textwidth}
\resizebox{\hsize}{!}{\includegraphics{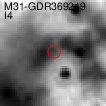}}
\end{minipage}
\begin{minipage}{0.02\textwidth}
  \phantom{.}
\end{minipage}
\begin{minipage}{0.23\textwidth}
\resizebox{\hsize}{!}{\includegraphics{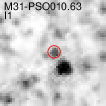}}
\end{minipage}
\begin{minipage}{0.23\textwidth}
\resizebox{\hsize}{!}{\includegraphics{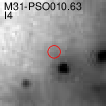}}
\end{minipage}

\caption{Cut-outs of about 30\arcsec\ x 30\arcsec\ (51x51 pixels of 0.60\arcsec) in the IRAC 1 and 4 filters
  centred on the CC. Cut levels are at the 0.5 and 99.5\% level.
  The red circle marks the nominal position and has a radius of 3 pixels, corresponding to approximately 1 FWHM of the point spread function.
  S Vul and GDR 404357 are plotted for comparison as MW stars not having an IR excess.
}
\label{Fig:IRAC}
\end{figure}

\end{appendix}

\end{document}